\documentclass[runningheads]{llncs}

\usepackage{times}

\usepackage{graphicx}
\usepackage{xcolor}

\usepackage[skip=0pt]{caption}
\usepackage{booktabs}
\usepackage{multirow}
\usepackage{numprint}
\npthousandsep{,}
\npdecimalsign{.}
\usepackage[inline]{enumitem}
\usepackage{balance}
\usepackage{amsmath}
\usepackage{amsfonts}
\usepackage{amssymb}
\usepackage{algorithm}
\usepackage{algpseudocode}
\usepackage{acronym}
\usepackage[numbers, square, sort&compress]{natbib}
\usepackage[export]{adjustbox}
\usepackage{arydshln}
\usepackage{bbm}
\usepackage{caption}
\usepackage{subcaption}
\usepackage{stackengine}
\usepackage{amsmath}
\usepackage[colorlinks=true,urlcolor=blue,citecolor=blue]{hyperref}

\newcommand{\headernodot}[1]{\vspace{1mm}\noindent\textbf{#1}}
\newcommand{\header}[1]{\headernodot{#1.}}

\acrodef{IR}{information retrieval}
\acrodef{CMR}{cross-modal retrieval}
\acrodef{i2t}{image-to-text}
\acrodef{t2i}{text-to-image}
\acrodef{VL}{vision-language}

\newcommand{\ourmodel}{Dense2Sparse}
\newcommand{\exact}{Exact@k}
\newcommand{\semantic}{Semantic@k}

\newcommand{\expnumber}[2]{${#1}\mathrm{e}{#2}$}
\newcommand{\negskip}{\vspace*{-1.5mm}}
\newcommand{\halfnegskip}{\vspace*{-0.75mm}}

\newcommand{\img}[1]{\mathbf{x}_{\mathcal{I}}^{#1}}
\newcommand{\capt}[2]{\mathbf{x}_{\mathcal{C}_{#2}}^{#1}}

\makeatletter
\newcommand{\printfnsymbol}[1]{%
  \textsuperscript{\@fnsymbol{#1}}%
}
\makeatother

\newcommand\sL{\ensuremath{\mathcal{L}}}

\newcommand\sR{\ensuremath{\mathcal{R}}}
\newcommand\sS{\ensuremath{\mathcal{S}}}

\newcommand\sX{\ensuremath{\mathcal{X}}}

\newcommand\sZ{\ensuremath{\mathcal{Z}}}

\newcommand\bx{\ensuremath{\mathbf{x}}}

\usepackage{color}

\author{
    Thong Nguyen\inst{1}\thanks{Co-first author.}
    \and
    Mariya Hendriksen\inst{2}\printfnsymbol{1}
    \and
    Andrew Yates\inst{1}
    \and
    Maarten de Rijke\inst{1}
    \institute{University of Amsterdam, The Netherlands
    \and
    AIRLab,  University of Amsterdam, The Netherlands
    \\
    \email{
    t.nguyen2@uva.nl, 
    m.hendriksen@uva.nl,\\
    a.c.yates@uva.nl,
    m.derijke@uva.nl}
    }
}

\authorrunning{T. Nguyen, M. Hendriksen, A. Yates, and M. de Rijke}

\begin{document}
\title{Multimodal Learned Sparse Retrieval with\\Probabilistic Expansion Control}

%

%
%
\maketitle              
\begin{abstract}
Learned sparse retrieval (LSR) is a family of neural methods that encode queries and documents into sparse lexical vectors that can be indexed and retrieved efficiently with an inverted index. We explore the application of LSR to the multi-modal domain, with a focus on text-image retrieval. While LSR has seen success in text retrieval, its application in multimodal retrieval remains underexplored. Current approaches like LexLIP and STAIR require complex multi-step training on massive data\-sets.  Our proposed approach efficiently transforms dense vectors from a frozen dense model into sparse lexical vectors. We address issues of high dimension co-activation and semantic deviation through a new training algorithm, using Bernoulli random variables to control query expansion. Experiments with two dense models (BLIP, ALBEF) and two datasets (MSCOCO, Flickr30k) show that our proposed algorithm effectively reduces co-activation and semantic deviation. Our best-performing sparsified model outperforms state-of-the-art text-image LSR models with a shorter training time and lower GPU memory requirements. Our approach offers an effective solution for training LSR retrieval models in multimodal settings. Our code and model checkpoints are available at 
\includegraphics[width=.75em]{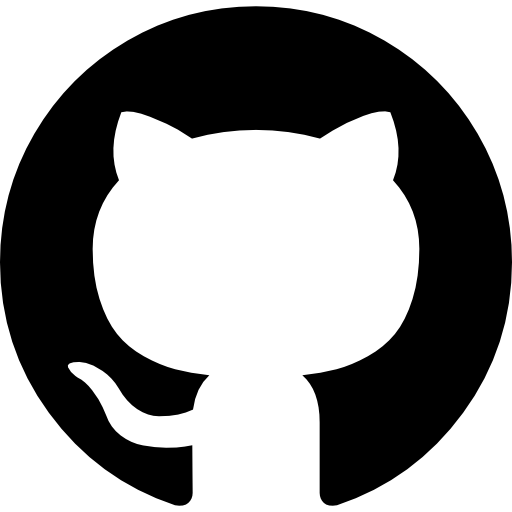} 
    \href{https://github.com/thongnt99/lsr-multimodal}{\nolinkurl{github.com/thongnt99/lsr-multimodal}}
\vspace{-1.5em}

\end{abstract}


\negskip\negskip
\section{Introduction}
\label{sec:introduction}
\negskip\negskip
Learned sparse retrieval (LSR)~\cite{formal2021splade, splade_pp, nguyen2023unified} typically employs transformer-based encoders to encode queries and documents into sparse lexical vectors (i.e., bags of weighted terms) that are compatible with traditional inverted index.
LSR has several nice properties. It provides an approach for effective and efficient neural retrieval, like dense retrieval, but with different advantages and trade-offs.
For example, sparse representations have the potential to be interpretable because they are aligned with a vocabulary, and they leverage inverted index software rather than approximate nearest neighbor search~\cite{nguyen2023unified}.
Empirically, LSR also shows advantages over single-vector dense models on retrieval generalization benchmarks~\cite{splade_pp, kamalloo2023resources}.

While LSR and dense retrieval are common in text retrieval, dense retrieval has taken the lead in multi-modal search. This is evident in state-of-the-art text-image pretraining methods like BLIP~\cite{li2022blip} and ALBEF~\cite{li2021align}, which rely on dense architectures. The preference for dense models arises because images, unlike text, consist of continuous pixel values, presenting a  challenge when they are mapped to discrete lexical terms. For multi-modal LSR, LexLIP~\cite{zhao2023lexlip} and STAIR~\cite{chen2023stair} are the only two recent methods that exhibit competitive results on standard benchmarks. However, both require complex multi-step training on extensive text-image pairs: LexLIP with up to 14.3 million pairs and STAIR with a massive 1 billion pairs, encompassing public and private data.

We approach the multi-modal LSR (MMLSR) problem by using a pre-trained dense model and training a small sparse projection head on top of dense vectors, using image-text dense scores as a supervision signal. Naively learning the projection layer leads to issues of
\begin{enumerate*}[label=(\roman*)]
\item high dimension co-activation and 
\item semantic deviation.
\end{enumerate*}
Issue~(i) happens when text and image sparse vectors excessively activate the same output dimensions, forming a sub-dense space inside the vocabulary space. 
Issue~(ii) means that produced output terms do not reflect the content of captions/images, making them not human-interpretable. To counter~(i) and~(ii), we propose a single-step training method with probabilistic term expansion control. By disabling term expansions, we force the projection to produce meaningful terms first, then gradually allow more term expansions to improve the effectiveness while also randomly reminding the model not to fully rely on expansion terms. This process is handled using Bernoulli random variables with a parameter scheduler to model the expansion likelihood at both caption and word levels.

Opting for dense to sparse projection, instead of training an MMLSR model from scratch, provides several advantages. First, it is aligned with the broader community effort to reduce the carbon footprint of training deep learning models \cite{luccioni2023counting}. By keeping the dense encoders frozen and learning a lightweight projection layer, we can avoid the double GPU training/inference cost of two models (dense \& sparse) while having more flexibility. Our approach enables the pre-computation of dense text and image vectors, allowing easy integration or removal of the projection layer based on available (dense or sparse) software and infrastructure. Moreover, this transformation may shed light on the interpretability of dense vectors, possibly contributing to a deeper understanding of the fundamental distinctions between these two multi-modal retrieval paradigms.

To understand the effectiveness and efficiency of the proposed training method, we conduct extensive experiments on two dense multi-modal models (BLIP, ALBEF) and two scene-centric~\cite{hendriksen2023scene} datasets (MSCOCO~\cite{lin2014microsoft}, Flickr30k~\cite{young2014image}).  We analyze the problems of dimension co-activation and semantic deviation under different settings. 

\header{Our contributions} The main contributions of our paper are: 
\begin{enumerate*}[label=(\roman*)]
    \item We propose a line of research for efficiently converting a multi-modal dense retrieval model to a multi-modal LSR model.
    \item We train a lightweight projection head to convert dense to sparse vectors and show that our sparsified models are faithful to dense models while outperforming previous multi-modal LSR models. The training is efficient and does not require ground-truth labels. 
    \item We identify the issues of high dimension co-activation and semantic deviation and propose a training method to address them.  
\end{enumerate*}

\negskip
\section{Related Work}
\label{sec:related-work}
\halfnegskip
\textbf{Learned sparse retrieval (LSR).} Learned sparse retrieval is a family of neural retrieval methods that encode queries and documents into sparse lexical vectors that can be indexed and searched efficiently with an inverted index. There are many LSR approaches in the literature on text retrieval~\cite{formal2021splade, zamani2018neural, nguyen2023adapting}; they are mainly built up from two types of encoder: MLP and MLM~\cite{nguyen2023unified}. The MLP encoder uses a linear feedforward layer placed on top of the transformers's last contextualized embeddings to predict the importance of input terms (similar to term-frequency in traditional lexical retrieval). The MLP encoder has no term expansion capability. On the other hand, the MLM encoder utilizes the logits of the masked language model (MLM) for weighting terms and selecting expansion terms. Splade~\cite{splade_pp, formal2021splade} is a recent state-of-the-art text-oriented LSR approach that employs the MLM encoder in both query and document side, while other methods \cite{macavaney2020expansion, lin2021few, dai2019context} use MLP encoders on both sides or only on the query side. Although it seems to be more beneficial to have expansion on both queries and documents, a recent study~\cite{nguyen2023unified} found that query and document expansion have a cancellation effect on text retrieval (i.e., having expansion on the document side reduces the usefulness of query expansion) and one could obtain near state-of-the-art results without query expansion. 

Unlike prior work focused on converting sparse to dense representations for hybrid ad-hoc text retrieval \citep{lin2021densifying, lin2023dense}, our work explores the reverse task of dense to sparse conversion in the multi-modal domain. This direction presents new challenges due to dimension co-activation and semantic deviation issues.
Ram et al. \cite{ram-etal-2023-token} interpreted text dense retrieval by zero-shot projection from dense to vocabulary space using a frozen MLM head. 
\citet{nguyen2023-hendriksen-multimodal} propose a simple sparse \ac{VL} bi-encoder without query expansion and evaluate the performance on the image suggestion task.
We aim for an efficient, effective, and semantically faithful drop-in sparse replacement of multi-modal dense retrieval, necessitating training of the projection layer.

\header{Cross-modal retrieval} \Ac{CMR} methods construct a multimodal representation space, where the similarity of concepts from different modalities can be measured using a distance metric such as a cosine or Euclidean distance. 
Some of the earliest approaches in \ac{CMR} utilized canonical correlation analysis~\cite{gong2014improving,klein2014fisher}. 
They were followed by a dual encoder architecture equipped with a recurrent and a convolutional component, the most prominent approaches in that area featured a hinge loss~\citep{frome2013devise, wang2016learning}. Later approaches further improved the effectiveness using hard-negative mining~\cite{faghri2017vsepp}. 

Later, the integration of attention mechanisms improved performance. This family of attention mechanisms includes dual attention~\cite{nam2017dual}, stacked cross-attention~\cite{lee2018stacked}, bidirectional focal attention~\cite{liu2019focus}. 
Another line of work proposes to use transformer encoders~\cite{vaswani2017attention} for this task~\cite{messina2021fine}, and adapts the BERT model~\cite{devlin2018bert} as a backbone~\citep{gao2020fashionbert, zhuge2021kaleido}.

A related line of work focuses on improving the performance on \ac{CMR} via modality-specific graphs~\cite{wang2021cross}, or image and text generation modules~\cite{gu2018look}. 
There is also more domain-specific work that focuses on \ac{CMR} in fashion~\citep{goei2021tackling, laenen2022cross}, e-commerce~\cite{hendriksen2022extending, hendriksen2022multimodal}, cultural heritage~\cite{sheng2021fine}, and cooking~\cite{wang2021cross}.



\negskip
\section{Background}
\label{sec:background}

\negskip
\textbf{Task definition.} We use the same notation as in previous work~\citep{zhang2020contrastive, brown2020smooth}. We work with a cross-modal dataset $\mathcal{D}$ that includes $N$ image-caption tuples: $\mathcal{D} = \left\{\left(\img{i}, \{\capt{i}{j}\}_{i=1}^{k} \right) \right\}_{i=1}^{N}$. Each tuple comprises an image $\img{}$ and $k$ associated captions $\{\capt{}{j}\}_{j=1}^{k}$.

The \acfi{CMR} task is defined analogously to the standard \acl{IR} task: given a query and a set of candidates, we rank all candidates w.r.t.\ their relevance to the query.
The query can be either a caption or an image.
Similarly, the set of candidate items can contain either images or captions.
\ac{CMR} is performed across modalities,  therefore, if the query is a caption then the set of candidates are images, and vice versa. 
Hence, the task comprises two subtasks:
\begin{enumerate*}[label=(\roman*)]
	\item \emph{caption-to-image retrieval}: retrieving images relevant to a caption query, and
	\item \emph{image-to-caption retrieval}: retrieving relevant captions that describe an image query.
\end{enumerate*} 
We focus on \emph{caption-to-image retrieval} only as it is more challenging, as reported by previous research~\cite{li2022blip, li2021align, zhao2023lexlip}.

\header{Sparsification-induced phenomena}
In this work, we investigate two phenomena arising during the sparsification process: dimension co-activation and semantic deviation.

\begin{definition}[Dimension co-activation]\rm
\label{def:dimension_coactivation}	
We define \emph{dimension co-activation} as sparse image and caption representations activating the same output dimensions, creating a sub-dense space within the vocabulary. While co-activation is essential for matching captions with images and can be measured by FLOPs, \emph{high co-activation} results in unnecessarily long posting lists, harming the efficiency of LSR. Establishing a clear threshold for \emph{high co-activation} is challenging, but we observe that beyond a certain point, increased FLOPs yield minimal improvements in effectiveness.
To quantify this effect, we use effectiveness metrics (e.g., R@k) in combination with the FLOPs metric:
\begin{equation}
\textstyle
    \mathrm{FLOPs} = \frac{1}{|\mathcal{C}||\mathcal{I}|}
    \sum_{\bx_{\mathcal{C}} \in \mathcal{C}}
    \sum_{\bx_{\mathcal{I}} \in \mathcal{I}}
    \mathbf{s}_{\mathcal{C}}^0 \cdot \mathbf{s}_{\mathcal{I}}^0
\end{equation}
where
$\mathcal{C}$ and $\mathcal{I}$ are caption and image collections, $\mathbf{s}_{\mathcal{C}}$, $\mathbf{s}_{\mathcal{I}}$ are sparse vectors of a caption $\bx_{\mathcal{C}}$ and an image $\bx_{\mathcal{I}}$.
\end{definition}
\begin{definition}[Semantic deviation]\rm
\label{def:semantic_deviation}	
We define \emph{semantic deviation} as the disparity between the semantic information in the visual or textual query and that in the sparse output terms. High co-activation suggests (but does not guarantee) semantic deviation.

Measuring semantic deviation directly is challenging, so we use two rough proxies, \emph{\exact} and \emph{\semantic}, defined as follows: 
\negskip
\begin{align}
\textstyle
    \exact &= \frac{1}{k} |\{t \mid t \in \bx_\mathcal{C},  t \in top_k(\mathbf{s}_{\mathcal{C}})\}|\\
\textstyle    
    \semantic &= \frac{1}{k} \sum_{\bx_{t}^{i} \in top_k(\mathbf{s}_{\mathcal{C}})}\max_{\bx_{t}^{j} \in \bx_\mathcal{C}}\frac{ f_{enc}(\bx_{t}^{i}) \cdot f_{enc}(\bx_{t}^{j}) }{\|f_{enc}(\bx_{t}^{i})\| \| f_{enc}(\bx_{t}^{j})\|}.
\end{align}
\emph{\exact} measures the ratio of overlapping terms between the input caption and the top-$k$ highest weighted output terms, providing a partial picture of semantic deviation without considering synonyms. $\semantic$ complements $\exact$ by calculating the averaged cosine similarity between static embeddings obtained using model $f_{enc}(\cdot)$ of top-$k$ output terms and input caption terms. Higher values for both metrics suggest less semantic deviation, implying better alignment of output terms with input captions.
\end{definition}






\section{Methodology}
\label{sec:methodology}
\negskip
\subsection{Model architecture}
\begin{figure}[t!]
    \centering
    \includegraphics[trim={8cm 7.5cm 8cm 6cm},clip, scale=0.6]{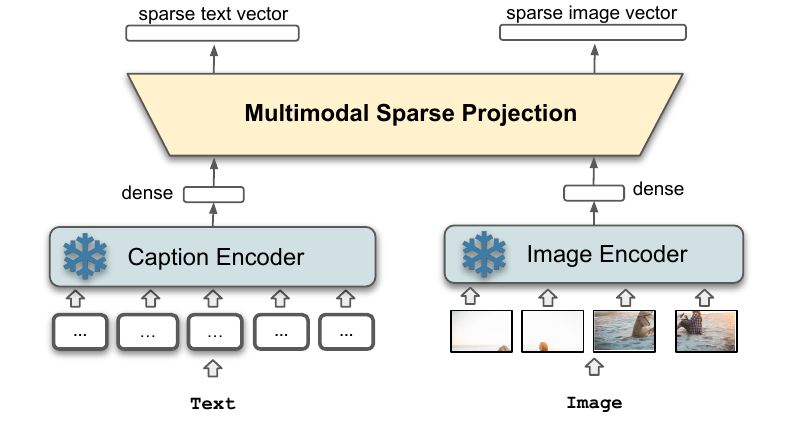}
    \vspace*{-2mm}
    \caption{The architecture of Dense2Sparse (D2S). The caption and image encoders are frozen, and the sparse projection is trained to project dense vectors to sparse vectors.} 
    \label{fig:dense2sparse}
\end{figure}

\negskip
The architecture of our \ourmodel{} model is visualized in Figure~\ref{fig:dense2sparse}. \ourmodel{} takes an image and a caption as input, projecting them into a $|V|$-dimensional space, where each dimension represents the weight of a corresponding vocabulary entry. The key components include two dense encoders, an image encoder $f_{\theta}^{\mathcal{I}}(\cdot)$ and a caption encoder $f_{\phi}^{\mathcal{C}}(\cdot)$, as well as a multimodal sparse projection head $g_{\psi}(\cdot)$.

\header{Dense image and text encoders}
The \emph{dense image encoder} $f_{\theta}^{\mathcal{I}}: \sX \rightarrow \sZ$ takes an input image $\mathbf{x}_{\mathcal{I}}$ and maps it into a latent space $\sZ = \sR^d$: $\mathbf{z}_{\mathcal{I}} =  f_{\theta}^{\mathcal{I}}(\bx_{\mathcal{I}})$, where $\mathbf{z}_{\mathcal{I}} \in \sR^d$. Similarly, the \emph{dense text encoder} $f_{\phi}^{\mathcal{C}}: \sX \rightarrow \sZ$ takes an input text (caption) $\bx_{\mathcal{C}}$, and maps it into a latent space $\sZ = \sR^d$: $\mathbf{z}_{\mathcal{C}} = f_{\phi}^{\mathcal{C}}(\bx_{\mathcal{C}})$, where $\mathbf{z}_{\mathcal{C}} \in \sR^d$. We obtain dense representations using BLIP and ALBEF as a backbone. Both encoders are frozen.

\header{Multimodal sparse projection head}
The \emph{multimodal sparse projection head} $g_{\psi}: \sZ \rightarrow \sS$ maps dense latent image and text representations into the sparse image and text vector space $\sS = \sR_{> 0}^{|V|}$:
\begin{align}
    \mathbf{s}_{\mathcal{C}} = g_{\psi}(\mathbf{z}_{\mathcal{C}}) \quad \text{and} \quad \mathbf{s}_{\mathcal{I}} = g_{\psi}(\mathbf{z}_{\mathcal{I}}).
\end{align}
The multimodal sparse projection head comprises four steps. First, we project the $d$-dimensional dense vector $\mathbf{z}$ to an $\omega$-dimensional dense vector: $\mathbf{z}_1 = \mathbf{W}_1 \mathbf{z}$, where $\mathbf{W}_1 \in \sR^{\omega \times d}$, $\mathbf{z} \in \sR^{d}$, and $\mathbf{z}_1 \in \sR^{\omega}$.
Second, we apply layer normalization:
\begin{align}
    \mathbf{z}_2 = \frac{ \mathbf{z}_1 - \mathbb{E}[\mathbf{z}_1] }{\sqrt{\mathit{Var}[\mathbf{z}_1] + \epsilon}}\cdot \gamma + \beta,
\end{align}
where $\mathbb{E}[\mathbf{z}_1]$ and $\mathit{Var}[\mathbf{z}_1]$ are the expectation and variance of $\mathbf{z}_1$, $\gamma$ and $\beta$ are learnable affine transformation parameters, and $\mathbf{z}_2 \in \sR^{\omega}$.
Third, we project $z_2$ to the vocabulary space $\sS = \sR_{> 0}^{|V|}$: $ \mathbf{s} = \mathbf{W}_{2} \mathbf{z}_2 $, where $\mathbf{W}_{2} \in \sR^{|V| \times \omega}$, $\mathbf{z}_2 \in \sR^{\omega}$, and $\mathbf{s} \in \sR^{|V|}$. $\mathbf{W}_{2}$ is initialized with vocabulary embeddings similar to the transformer-masked language model.
Fourth, we remove negative weights and apply a logarithmic transformation to the positive weights: $\mathbf{s} = \log_{e}(1+ \max(0, \mathbf{s}))$, where $\mathbf{s} \in \sR_{>0}^{|V|}$. The resulting $|V|$-dimensional sparse vector is aligned with the vocabulary, and each dimension represents the weight of the corresponding vocabulary entry. This projection head is similar to the MLM head employed in previous work~\cite{splade_pp, macavaney2020expansion}. 

\begin{algorithm}[t!]
\caption{Multimodal LSR training with probabilistic expansion control}
\label{alg:lsr_training}
\mbox{}\vspace*{-5mm}
\begin{multicols}{2}
\begin{algorithmic}
\State {\bfseries Input: }
image-caption pair ($\bx_{\mathcal{I}}$, $\bx_{\mathcal{C}}$),
caption encoder $f_{\phi}^{\mathcal{C}}$,
image encoder $f_{\theta}^{\mathcal{I}}$,
sparse projection head $g_{\psi}$,
loss function $\sL$, and
expansion rate function $f_{\text{incr}}$. 
\State
\State \texttt{$p^v_i$} $\gets 1 - df^v_i$
\State \texttt{$p_c$} $\gets 0$
\State
\For{\texttt{epoch}}
    \For{\texttt{batch}}
        \State $\mathbf{z}_{\mathcal{C}} \gets f_{\phi}^{\mathcal{C}}(\bx_{\mathcal{C}})$, \text{ } $\mathbf{z}_{\mathcal{I}} \gets f_{\theta}^{\mathcal{I}}(\bx_{\mathcal{I}})$
        \State $\mathbf{s}_{\mathcal{C}} \gets g_{\psi}( \mathbf{z}_{\mathcal{C}} )$, \text{ } $\mathbf{s}_{\mathcal{I}} \gets g_{\psi}( \mathbf{z}_{\mathcal{I}} )$
        \State \texttt{$\mathcal{E}_{\mathcal{C}}$} $\sim$ \text{Ber}(\texttt{$p_c$}), \text{ } \texttt{$\mathcal{E}^v_i$} $\sim$ \text{Ber}(\texttt{$p^v_i$})
        \State $\mathbf{\overline{s}}_{\mathcal{C}} \gets \mathrm{EXPAND}(\bx_{\mathcal{C}}, \mathbf{s}_{\mathcal{C}}, \mathcal{E}_{\mathcal{C}}, \mathcal{E}^v_i) $
        \State  $\sL \gets \sL(\mathbf{\overline{s}}_{\mathcal{C}}, \mathbf{s}_{\mathcal{I}}, \mathbf{z}_{\mathcal{I}}, \mathbf{z}_{\mathcal{C}})$ 
    \EndFor
    \State \texttt{$p_c$} $\gets$ $f_{incr}$(\texttt{$p_c$}), \texttt{ $p^v_i$} $\gets$ $f_{incr}$(\texttt{$p^v_i$})
\EndFor
\end{algorithmic}
\columnbreak
\begin{algorithmic}
\Function{EXPAND}{$\bx_{\mathcal{C}}$, $\mathbf{s}_{\mathcal{C}}$, $\mathcal{E}_{\mathcal{C}}$, $\mathcal{E}^v_i$}
    \For{$0 \leq i < \text{batch\_size}$}
        \For {$0 \leq k < |V|$}
            \If{$v_k \notin \bx_{\mathcal{C}}$}
                \State ${\mathbf{s}_{\mathcal{C}}}_{i,k} \gets {\mathbf{s}_{\mathcal{C}}}_{i,k} \cdot \mathcal{E}_{\mathcal{C}} \cdot e^v_k$
            \Else 
                \State ${\mathbf{s}_{\mathcal{C}}}_{i,k} \gets {\mathbf{s}_{\mathcal{C}}}_{i,k} \cdot \mathcal{E}^v_k$
            \EndIf
        \EndFor 
    \EndFor
    \State \Return $ \mathbf{s}_{\mathcal{C}}$
\EndFunction
\end{algorithmic}
\end{multicols}
\vspace*{-3mm}
\end{algorithm}

\header{Probabilistic expansion control}
Without any intervention, training the projection module with a standard contrastive loss could lead to high-dimension co-activation and semantic deviation as defined previously. This phenomenon affects the efficiency of an inverted index and the interpretability of the outputs. To mitigate this problem, we propose a single-step training algorithm with probabilistic lexical expansion control. It is described in Algorithm~\ref{alg:lsr_training}.

We define a Bernoulli random variable $\mathcal{E} \sim \mathit{Ber}(p),\ p \in [0, 1]$ and use it to control textual query expansion.
We consider a caption-level and a word-level expansion. The \emph{caption-level expansion} is controlled by the random variable $\mathcal{E}_{\mathcal{C}} \sim \mathit{Ber}(p_{\mathcal{C}})$. If $\mathcal{E}_{\mathcal{C}}= 1$ the expansion is allowed, while $\mathcal{E}_{\mathcal{C}}= 0$ means the expansion is not allowed. 
Analogously, the \emph{word-level expansion}, or the expansion to the $i$-th word in the vocabulary, is regulated by the random variable $\mathcal{E}^v_i \sim \mathit{Ber}(p^v_i)$.

The parameters $p_{\mathcal{C}}$ and $p^v_i$ define the likelihood of caption-level and word-level expansion within a given training epoch. 
During training, we initially set the caption-level expansion probability, $p_{\mathcal{C}}$, to zero. This initial value prevents the expansion of textual queries, forcing the model to project images onto relevant tokens belonging to the captions they were paired with.
This approach facilitates the meaningful projection of dense vectors onto relevant words in the vocabulary. However, it adversely impacts retrieval effectiveness, as the model cannot expand queries. As a consequence, the model's ability to handle semantic matching is limited.
To gradually relax this constraint, we use a scheduler that incrementally increases the value of $p$ after each epoch until it reaches a maximum value of one in the final epoch.
In each epoch, we sample the values of $\mathcal{E}$ per batch and enforce expansion terms to be zero when $\mathcal{E}_{\mathcal{C}}$ equals zero.
Similarly, for word-level expansion, we initialize the expansion probability of the $i$-th word $p^v_i$ to $1 - df^v_i$ where $df^v_i$ is the normalized document frequency of vocabulary element $v_i$ in the caption collection $\mathcal{C}$.
This setting discourages the expansion of more frequent terms because they are less meaningful and can hinder the efficiency of query processing algorithms.
We relax each $p^v_i$ after every epoch, ensuring that it reaches a maximum value of one at the conclusion of the training process.
The expansion rate increase after each epoch is defined as follows:
\begin{equation}
f_{\text{incr}}(p) = 
\begin{cases}
    p + \frac{1}{\textit{\# epochs}}, & \text{for caption-level expansion} \\
    p + \frac{df^v_i}{\textit{\# epochs}}, & \text{for word-level expansion}.
\end{cases}
\end{equation}



\negskip
\subsection{Training loss} We train our \ourmodel{} using a loss that represents a weighted sum of a bidirectional loss and a sparse regularization parameter. The bidirectional loss is based on the following one-directional loss:
\begin{align}
    \notag
    \ell^{(\mathcal{A} \rightarrow \mathcal{B})} &= 
    - \left(\frac{\exp(\mathbf{z}_{\mathcal{A}}^{\intercal}\mathbf{z}_{\mathcal{B}} / \tau )}{\sum_{\mathcal{I*}} \exp(\mathbf{z}_{\mathcal{A}}^{\intercal}\mathbf{z}_{\mathcal{I*}} / \tau )}\right)
    \log_2\left(\mathrm{SoftMax}[\mathbf{s}_{\mathcal{A}}^{^\intercal}\mathbf{s_{\mathcal{B}}}]\right),
\end{align}
%
where
$\mathbf{s_{\mathcal{A}}} \in \sR^{|V|}_{>0}$ and $\mathbf{s_{\mathcal{B}}} \in \sR^{|V|}_{>0}$ are sparse vectors, $\mathbf{z_{\mathcal{A}}} \in \sR^{d}$ and $\mathbf{z_{\mathcal{B}}} \in \sR^{d}$ are dense vectors, and $\tau \in \sR_{>0}$ is a temperature parameter.

The resulting loss is formalized to capture both bidirectional losses and sparse regularization. The overall loss $\sL$ is defined as: 
\begin{align}
    \label{eq:contrastive_loss}
    \sL &=
    (1 - \lambda)
    \underbrace{[\ell^{(\mathcal{I} \rightarrow \mathcal{C})} + \ell^{(\mathcal{C} \rightarrow \mathcal{I})}]}_{\textit{bidirectional loss}}
    +
    \lambda \underbrace{ \eta [L_1(\mathbf{s}_{\mathcal{I}}) + L_1(\mathbf{s}_{\mathcal{C}})]}_{\textit{sparse regularization parameter}},
\end{align}
where $\ell^{(\mathcal{I} \rightarrow \mathcal{C})}$ is an image-to-caption loss, $\ell^{(\mathcal{C} \rightarrow \mathcal{I})}$ is a caption-to-image loss;
$\lambda = [0,1]$ is a scalar weight, $\eta = [0,1]$ is a sparsity regularization parameter, and $L_1(\mathbf{x}) = \|\mathbf{x}\|_1$ is $L_1$ regularization. It is worth noting that the loss utilizes dense scores for supervision, a strategy found to be more effective than using ground truth labels. 

\negskip\negskip
\section{Experiments and Results}
\label{sec:experiments}
\negskip
\subsection{Experimental setup}
\negskip
\textbf{Datasets.} We trained and evaluated our models on two widely used datasets for text-image retrieval: MSCOCO~\cite{lin2014microsoft} and Flickr30k~\cite{plummer2015flickr30k}. Each image in the two datasets is paired with five short captions (with some exceptions). We re-used the splits from~\cite{karpathy2015deep} for training, evaluating, and testing. The splits on MSCOCO have 113.2k pairs for training, and 5k pairs for each validation/test set. Flickr30 is smaller with 29.8k/1k/1k for train, validation, test splits respectively. The best model is selected based on the validation set and evaluated on the test set. 

\header{Evaluation metrics} To evaluate model performance and effectiveness, we report R@k where $k = \{1, 5\}$, and MRR@10 using the \textit{ir\_measures}~\cite{macavaney2022streamlining} library.

\header{Implementation and training details} The caption and image dense vectors of BLIP \cite{li2022blip} and ALBEF \cite{li2021align} models are pre-computed with checkpoints from the larvis library~\cite{li2022lavis}. We train our models to convert from dense vectors to sparse vectors on a single A100 GPU with a batch size of 512 for 200 epochs. The training takes around 2 hours and only uses up to around 10 GB of GPU memory. We set the temperature $\tau$ to $0.001$ and experiment with sparse regularization weights $\eta \in [1e-5, 1e-2]$.

\negskip
\subsection{Results and discussion}

\begin{figure}[t!]
     \centering
     \begin{subfigure}[b]{0.45\textwidth}
         \centering
         \includegraphics[width=\textwidth]{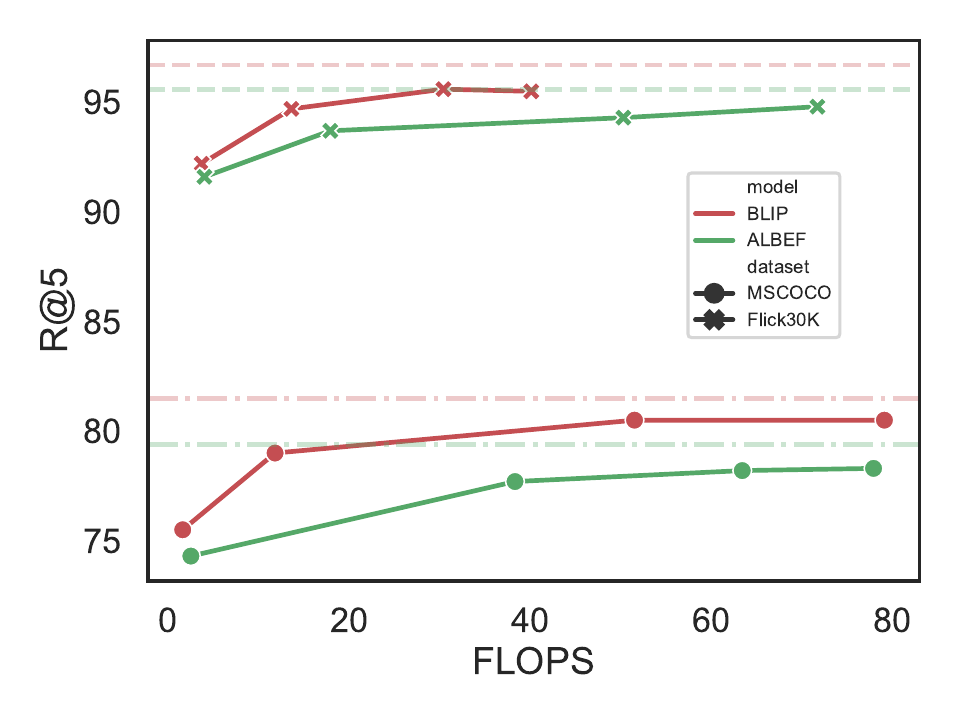}
         \vspace*{-5mm}
         \caption{Efficiency vs. effectiveness of sparsified models}
         \label{fig:lsr_efficiency_effectiveness}
     \end{subfigure}
     \hfill
     \begin{subfigure}[b]{0.45\textwidth}
         \centering
         \includegraphics[width=\textwidth]{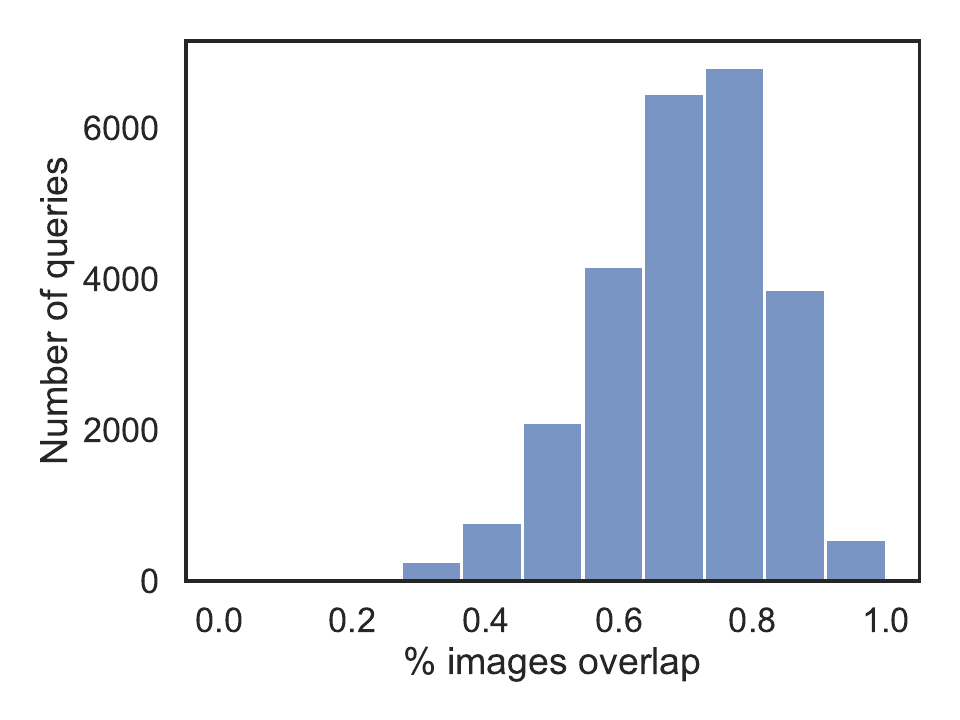}
         \vspace*{-5mm}
         \caption{Fraction of overlapping images in the top-10 by sparsified and dense BLIP model.}
         \label{fig:images_overlap}
     \end{subfigure}
     \caption{Sparisified models compared to original dense models.}
\end{figure}

\negskip
\noindent%
\textbf{RQ1: How effective and efficient is the proposed method for converting dense to sparse?}
We trained various \ourmodel{} models (D2S) using our proposed training method with different sparse regularization weights ranging from $1e-5$ to $1e-2$. Figure \ref{fig:lsr_efficiency_effectiveness} illustrates the effectiveness and efficiency of these variations, with detailed results presented in Table \ref{tab:dense2sparse_results}.
\begin{table}[t!]
    \caption{The effectiveness of sparsified models (D2S) and baselines.  ($^\dagger p<0.05$ \textit{with paired two-tailed t-test comparing D2S to the dense model with Bonferroni correction})}
    \label{tab:dense2sparse_results}
    \centering
    \resizebox{\textwidth}{!}{  
    \begin{tabular}{@{\extracolsep{2pt}}lcccccccc@{}}
    \toprule
    \multirow{2}{*}{\textbf{Model}} & \multicolumn{4}{c}{\textbf{MSCOCO 
 (5k)}}  & \multicolumn{4}{c}{\textbf{Flickr30k (1k)}} \\  
    \cmidrule{2-5} \cmidrule{6-9}
    & R@1$\uparrow$ & R@5$\uparrow$ & MRR@10$\uparrow$ & FLOPs$\downarrow$ &
    R@1$\uparrow$ & R@5$\uparrow$ & MRR@10$\uparrow$ & FLOPs$\downarrow$\\ 
    \midrule
    \multicolumn{7}{l}{\textit{T2I Dense Retrieval}}\\
    \midrule
    COOKIE \cite{DBLP:conf/iccv/WenXHLXS21} & 46.6 & 75.2 & - & - &  68.3 & 91.1 & - & - \\ 
    COTS (5.3M) \cite{DBLP:conf/cvpr/LuFHG0W22} & 50.5 & 77.6 & - & - &  75.2 & 93.6 & - & - \\
    ALBEF \cite{li2021align} &  53.1 & 79.3 & 64.3 & -  &  79.1 & 94.9 & 86.6 & - \\
    BLIP \cite{li2022blip} & \textbf{57.3} & \textbf{81.8} & \textbf{67.8} & - & \textbf{83.2} & \textbf{96.7} & \textbf{89.3} & - \\ 
    \midrule
    \multicolumn{7}{l}{\textit{T2I Sparse Retrieval}}\\
    \midrule
    VisualSparta & 45.1 & 73.0 & - & -  & 57.1 & 82.6 & - & - \\
    STAIR (zero-shot) & 41.1 & 56.4 & - & - & 66.6 & 88.7 & - & - \\ 
    LexLIP (4.3M) & 51.9 &	78.3 & - & - & 76.7 & 93.7 & - & -\\ 
    LexLIP (14.3M) & 53.2 &  79.1 & - & - & 78.4 & 94.6 &  - & -  \\ 
    \midrule
    D2S (ALBEF, $\eta=1e-3$) & 49.6$^\dagger$ & 77.7$^\dagger$ & 61.4$^\dagger$ & 18.7 & 74.2$^\dagger$ & 93.8$^\dagger$ & 82.6$^\dagger$ & 21.7\\ 
    D2S (ALBEF, $\eta=1e-5$) & 50.7$^\dagger$ & 78.2$^\dagger$ & 62.4$^\dagger$ & 74.2 & 75.4$^\dagger$ & 94.3$^\dagger$ & 83.6$^\dagger$ & 64.3\\ 
    D2S (BLIP, $\eta=1e-3$) & 51.8$^\dagger$ & 79.3$^\dagger$ & 63.4$^\dagger$ & \textbf{11.5} & 77.1$^\dagger$	 & 94.6$^\dagger$ & 	84.6$^\dagger$ & \textbf{\phantom{0}9.9}\\ 
    D2S (BLIP, $\eta=1e-5$) & \textbf{54.5}$^\dagger$ & \textbf{80.6}$^\dagger$ & \textbf{65.6}$^\dagger$ & 78.4 & \textbf{79.8}$^\dagger$ & \textbf{95.9}$^\dagger$ & \textbf{86.7}$^\dagger$ & 39.5 \\
    \bottomrule
    \end{tabular}
    } 
\end{table}
Firstly, we observe that increasing the sparse regularization weight enhances model efficiency (reduced FLOPs) but reduces its effectiveness (lower Recall and MRR). On the MSCOCO dataset, our most efficient sparse BLIP model ($\eta=1e-2$) achieves a R@1 of $47.2$ and MRR@10 of $58.5$ with the lowest FLOPs value of $1.6$. Relaxing the regularization weight to $1e-3$ results in an approximately $10\%$ increase in R@1 to $51.8$ and a similar rise in MRR@10 to $63.4$, albeit at the expense of around $7$ times higher FLOPs (less efficient).

Further relaxing the sparse regularization gradually brings the sparsified model's effectiveness closer to the original dense model, while reducing the efficiency. The most effective sparsified BLIP model with $\eta=1e-5$ performs competitively with the original dense version ($54.5$ vs. $57.3$) and outperforms other dense baselines. 

Additionally, we observe a diminishing gap between dense and sparsified models as we assess recalls at higher cutoff positions, such as $R@5$ and $R@10$. Similar trends are observed across different datasets, including Flickr30k and MSCOCO, as well as among different dense models, including BLIP and ALBEF. This indicates the broad applicability of our proposed approach to diverse datasets and models.

\headernodot{RQ2: How does our sparsified model compare to state-of-the-art multi-modal LSR models?}
In this research question, we compare our sparsified models with existing LSR baselines, namely Visual Sparta, STAIR, and LexLIP. Currently, neither the code nor the checkpoints for these baselines are publicly available. Therefore, we rely on the numbers reported in their respective papers for comparison, excluding the FLOPs.

STAIR and LexLIP are two of the most recent multimodal LSR approaches, both trained on large datasets, with STAIR utilizing 1 billion internal text-image pairs. In contrast, our proposed method leverages pretrained dense retrieval models to efficiently learn a lightweight sparse projection for converting dense vectors to sparse vectors. 

The effectiveness of our methods and the baselines on MSCOCO and Flickr30k is presented in Table \ref{tab:dense2sparse_results}. Notably, our efficient model, D2S(BLIP, $\eta=1e-3$), performs competitively with LexLIP trained on 4.3 million text-image pairs at R@1. Its R@5 is slightly better than LexLIP (4.3M) and comparable to the LexLIP model trained on 14.3 million pairs. With a lower sparse regularization, our D2S(BLIP, $\eta$=\expnumber{1}{-5}) model significantly outperforms all baselines on both MSCOCO and Flickr30k. On MSCOCO, its R@1 is 21\%, 5\%, and 2.8\% higher than the R@1 of Visual Sparta, LexLIP (4.3M), and LexLIP (14.3M), respectively. All our models outperform Visual Sparta and STAIR, although this comparison with STAIR uses a zero-shot setting, because we lack access to their code and checkpoints for fine-tuning STAIR further with in-domain data. 

We kept the dense encoders frozen, so the effectiveness of our sparsified models is inherently bounded by the dense results. Our sparsified ALBEF models, for example, exhibit slightly lower overall effectiveness since their corresponding dense performance is lower than that of BLIP's dense scores. Nonetheless, our sparsified ALBEF models are also comparable with LexLIP variants.

\headernodot{RQ3: Does the proposed training method help address the dimension co-activation and semantic deviation issues?}
As discussed in Section \ref{sec:background}, high co-activation increases posting list length, impacting inverted index efficiency. We examine this impact by analyzing FLOPs alongside model effectiveness metrics. Table \ref{tab:dense2sparse_results} presents results for models trained with our method and three baseline variants, with fixed expansion rates of $0$ and $1$ in the first two baselines. The third baseline ($exp=c$) explores the influence of word-level expansion control, excluding it from our training method.

At an expansion rate of zero, models project the caption's dense vector only onto terms from the caption, with all other projections forced to zero. The image projector must then learn to align the image vector with terms in the paired captions.
Conversely, setting $exp$ to $1$ gives the model the freedom to project onto any output vectors, making it more inclined toward dimension co-activation.

\begin{table}[t!]  
    \caption{The dimension co-activation effect of Dense2Sparse (D2S) variations.}
    \label{tab:densification}
    \centering
    \resizebox{\textwidth}{!}{  
    \begin{tabular}{@{\extracolsep{2pt}}lccclcccl@{}}
    \toprule
    \multirow{2}{*}{\textbf{Model} (D2S variations)} & \multicolumn{4}{c}{\textbf{MSCOCO 
 (5k)}}  & \multicolumn{4}{c}{\textbf{Flickr30k (1k)}} \\  
    \cmidrule{2-5} \cmidrule{6-9}
    & R@1$\uparrow$ & R@5$\uparrow$ &  MRR@10$\uparrow$ & FLOPs$\downarrow$ &  R@1$\uparrow$ & R@5$\uparrow$ & MRR@10$\uparrow$  & FLOPs$\downarrow$\\ 
    \midrule
    (BLIP, $\eta=1e-3$, $exp=0$) & 45.5 & 73.0 & 57.3 & \phantom{00}2.8 & 68.9 & 89.5 & 77.8 & \phantom{00}3.0 \\ 
    (BLIP, $\eta=1e-3$,  $exp=1$) & 53.4 & 80.0 & 64.6 & \phantom{0}49.1 & 79.5 & 95.5 & 86.4 & \phantom{0}50.3 \\ 
    (BLIP, $\eta=1e-3$, $exp=c$) & 51.9 & 79.0 & 63.4 & \phantom{0}11.8 & 77.3 & 94.7 & 84.8 & \phantom{0}13.6 \\ 
    (BLIP, $\eta=1e-3$, $exp=c+w$) & 51.8& 79.3 & 63.4 & \phantom{0}11.5	& 77.1 & 94.6 & 84.6 & \phantom{00}9.9\\
    \midrule 
    (BLIP, $\eta=1e-5$, $exp=0$) & 47.2 & 74.4 & 58.8 & \phantom{00}3.2 & 72.3 & 91.8 & 80.7 & \phantom{00}3.5 \\ 
    (BLIP, $\eta=1e-5$, $exp=1$) & 55.9 & 81.3 & 66.8 & 343 & 81.4 & 96.0 & 87.7 & 213 \\ 
    (BLIP, $\eta=1e-5$, $exp=c$) & 54.7 & 80.5 & 65.8 & \phantom{0}79.1 & 79.9 & 95.5 & 86.7 & \phantom{0}40.1   \\
    (BLIP, $\eta=1e-5$, $exp=c+w$) & 54.5 & 80.6 & 65.6	 & \phantom{0}78.4 & 79.8 & 95.9 & 86.7 & \phantom{0}39.5 \\ 
    \midrule
    (ALBEF, $\eta=1e-3$, $exp=0$) & 43.8 & 71.8 & 55.7 & \phantom{00}2.5 &  65.8 & 88.3 & 75.4 & \phantom{00}3.0\\ 
    (ALBEF, $\eta=1e-3$, $exp=1$) & 50.9 & 78.4 & 62.5 & \phantom{0}68.2 & 75.7 & 94.2 & 83.8 & \phantom{0}61.9 \\ 
    (ALBEF, $\eta=1e-3$, $exp=c$) & 49.7 & 77.7 & 61.5 & \phantom{0}38.3 & 74.6 & 93.7 & 82.8 & \phantom{0}17.9\\ 
    (ALBEF, $\eta=1e-3$, $exp=c+w$) & 49.6 & 77.7 & 61.4 & \phantom{0}18.7 & 74.2 & 93.8 & 82.6 & \phantom{0}21.7\\ 
    \midrule 
    (ALBEF, $\eta=1e-5$, $exp=0$) & 45.9 & 73.9 & 83.0 & \phantom{00}3.4 & 68.1 & 90.0 & 77.6 & \phantom{00}3.2 \\ 
    (ALBEF, $\eta=1e-5$, $exp=1$) &  52.4 & 78.7 & 63.7 & 283 & 77.2 & 94.6 & 84.8 & 210  \\ 
    (ALBEF, $\eta=1e-5$, $exp=c$) & 51.2 & 78.3 & 62.8 & \phantom{0}77.9  & 76.4 & 94.8 & 84.0 & \phantom{0}71.7 \\ 
    (ALBEF, $\eta=1e-5$, $exp=c+w$) & 50.7 & 78.2 & 62.4 & \phantom{0}74.2 & 75.4 & 94.3 & 83.6 & \phantom{0}64.3 \\ 
    \bottomrule
    \end{tabular}
    } 
\end{table}

In Table \ref{tab:densification}, rows with ($exp=0$) show models with no expansion, resulting in remarkably low FLOPs, with each query averaging $2$ to $3$ overlapping terms with each document. However, disabling expansion reduces the model's ability for semantic matching, leading to modest effectiveness (45--47R@1 on MSCOCO and 68--72R@1 on Flickr30k with varying sparsity).
%
Enabling non-regulated expansion ($exp=1$) significantly improves model effectiveness (50--55 R@1 on MSCOCO and 75--79R@1 on Flickr30k with various regularization weights). However, this improvement comes at the cost of substantially increased FLOP scores, sometimes by up to 100 times, making sparsified vectors very computationally expensive. Ultimately, the resulting models behave like dense models, which is an undesired effect.

Our training method, which incorporates expansion control at the caption and word levels, is designed to gradually transition from one extreme ($exp=0$) to the other ($exp=1$). During training, we allow a likelihood of expansion, which increases progressively to over time. However, we also introduce random elements, represented by a random variable, to remind the model to remain faithful to the original captions/images.

The results, displayed in rows labeled with $exp=c+w$, demonstrate that our approach strikes a better balance between efficiency and effectiveness. It achieves competitive levels of effectiveness compared to models with $exp=1$ while requiring only half or a third of the computational operations (FLOPs). For example, on MSCOCO with the BLIP model, Dense2Sparse ($\eta=1e-3$) achieves a performance of $51.8$ R@1 (compared to $53.4$ when $exp=1$) with just $11.8$ FLOPs, making it four times more efficient than the $exp=1$ baseline. With the same setting, our method achieves 14\% higher R@1 and 11\% higher MRR@10 than the baseline with no expansion ($exp=0$). Compared to the baseline without word-level expansion control, no significant differences are observed in terms of efficiency and effectiveness. Thus, caption-level expansion control alone seems sufficient for achieving reasonable efficiency and effectiveness. Similar results are noted across various settings, datasets, and dense models.

\begin{table}[t!]  
    \caption{Semantic deviation on different Dense2Sparse (D2S) variations. ($^\dagger p<0.01$ \textit{with paired two-tailed t-test comparing exp=c to exp=1})}
    \label{tab:semantic_deviation}
    \centering
    \begin{tabular}{@{\extracolsep{2pt}}lcccc@{}}
    \toprule
    \multirow{2}{*}{\textbf{Model} (D2S variations)} & \multicolumn{2}{c}{\textbf{MSCOCO 
 (5k)}}  & \multicolumn{2}{c}{\textbf{Flickr30k (1k)}} \\  
    \cmidrule{2-3} \cmidrule{4-5}
    & Exact@20 & Semantic@20 & Exact@20 & Semantic@20 \\
    \midrule
    (BLIP, $\eta=1e-5$, $exp=c$) & 20.0$^\dagger$ & 60.1$^\dagger$ & 18.3$^\dagger$	& 58.0$^\dagger$ \\
    (BLIP, $\eta=1e-5$, $exp=1$) & \phantom{0}6.9	& 48.5 & \phantom{0}3.2 & 40.7\\
    \midrule
    (BLIP, $\eta=1e-3$, $exp=c$) & 25.0$^\dagger$ & 63.2$^\dagger$ & 23.1$^\dagger$ & 60.6$^\dagger$ \\
    (BLIP, $\eta=1e-3$, $exp=1$) & \phantom{0}2.5 & 42.0 & \phantom{0}2.2 & 41.1\\
    \midrule 
    (ALBEF, $\eta=1e-5$, $exp=c$) & 20.5$^\dagger$ & 61.0$^\dagger$ &	19.2$^\dagger$ &	59.8$^\dagger$ \\
    (ALBEF, $\eta=1e-5$, $exp=1$) & \phantom{0}5.6 & 43.5 & 	\phantom{0}1.2 &	40.5 \\ 
    \midrule 
    (ALBEF, $\eta=1e-3$, $exp=c$) & 15.1$^\dagger$ &	51.3$^\dagger$ & 19.6$^\dagger$ &	56.4$^\dagger$ \\
    (ALBEF, $\eta=1e-3$, $exp=1$) & \phantom{0}1.6 & 40.6 & \phantom{0}1.3 & 41.5 \\
    \bottomrule
    \end{tabular}
\end{table}

Sparse representations contain interpretable output dimensions aligned with a vocabulary. However, training a D2S model without our expansion regulation leads to semantic deviation, turning vocabulary terms into non-interpretable latent dimensions. We assess this effect using \exact{} and \semantic{} metrics (defined in Section \ref{sec:background}), reporting results in Table \ref{tab:semantic_deviation} and providing qualitative examples in Table \ref{tab:semantic_deviation_examples}.
\begin{table}[t!]
    \caption{Examples of semantic deviation. We show the top-10 terms per model. }
    \label{tab:semantic_deviation_examples}    \centering
    \resizebox{\linewidth}{!}{
    \begin{tabular}{@{} m{0.324\linewidth}@{\hspace{.74em}}m{0.374\linewidth}@{\hspace{.74em}} m{0.3\linewidth}@{}}
    \toprule
    \textbf{Caption, Image} & \textbf{D2S ($\eta=1e-3$, exp=c)} & \textbf{D2S ($\eta=1e-3$, exp=1.0)} \\ 
    \midrule
    A man with a red helmet on a small moped on a dirt road     & dirt, mo, motor, motorcycle, bike, red, riding, features, soldier, \#\#oot  & ,  accent  "  yourself  natural  may  while  officer  english  ac\\
    \hdashline
    \addlinespace[1ex]
    \centering \raisebox{-0.6\height}{\includegraphics[width=1.5cm, valign=m]{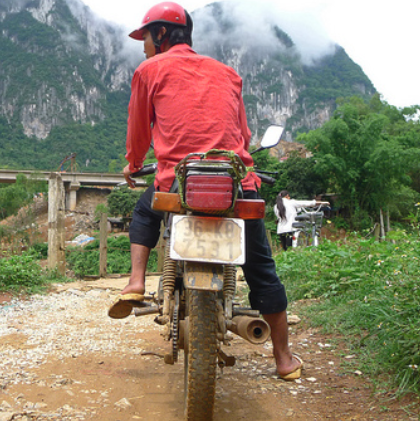}}  
 & mountain mountains bike bee dirt mo red path \#\#oot person man riding bicycle & accent  ship  natural  de  crown  yourself  "  ra  now  wild \\ 
    \midrule
    A women smiling really big while holding a Wii remote. & lady  woman  smile  women  remote  laughing  wii  smiling  video  controller &  ,  kai  called  forces  rush  lee  war  oil  like  \#\#h \\
    \hdashline
    \addlinespace[1ex]
    \centering \raisebox{-0.6\height}{\includegraphics[width=1.5cm, valign=m]{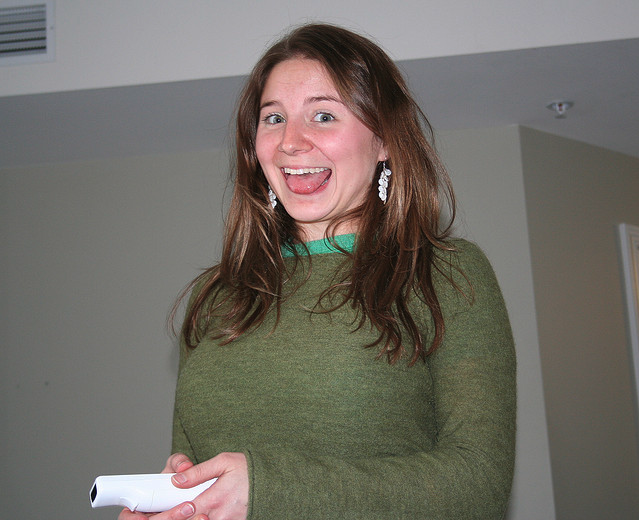}} &  smile  after  green  woman  smiling  sweater  remote  lady  wii  her & tall  kai  forces  oil  rush  met  war  college  thus  there 
    \\ 
    \midrule
    A couple of dogs sitting in the front seats of a car. & dogs  dog  car  backseat  seat  couple  vehicle  sitting  two  puppy & ,  electric  stood  forest  national  master  help  arts  fc  -\\ 
    \hdashline
    \addlinespace[1ex]
    \centering \raisebox{-0.6\height}{\includegraphics[width=1.5cm, valign=m]{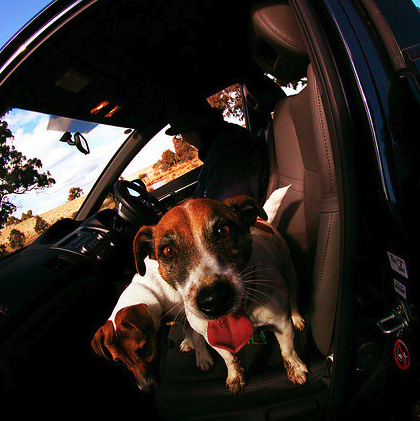}} & dog  car  dogs  puppy  out  vehicle  pup  inside  early  open & stood  forest  national  electric  master  twice  grant  men  para  yet\\ 
    \bottomrule
    \end{tabular}}
\end{table}

Uncontrolled models (with $exp=1$) exhibit lower Exact@20 and Semantic@20 than our expansion-controlled models ($exp=c$). In the top 20 terms of uncontrolled models, only one or none are in the original captions, while controlled models generate $3$ to $5$ caption terms. The low Semantic@20 of the uncontrolled models also suggests low relatedness of output terms to the caption terms. This implication could be further supported by the examples demonstrated in Table \ref{tab:semantic_deviation_examples}. Uncontrolled models generate random terms, while our method produces terms that more faithfully reflect captions and images. Most top-10 terms from our method are relevant to the input, including a mix of original terms and synonyms (e.g., ``dog'' vs.\ ``puppy'', ``car'' vs.\ ``vehicle'').

\headernodot{RQ4: Is the sparsified model faithful to the dense model?}
\begin{table}[t!]
    \caption{Correlation between dense and different variations of Dense2Sparse (D2S).}
    \label{tab:dense-sparse-correlation}
    \centering
    \begin{tabular}{@{\extracolsep{2pt}}lcccccc@{}}
    \toprule
    \multirow{2}{*}{\textbf{Model} (D2S variations)} & \multicolumn{3}{c}{\textbf{MSCOCO 
 (5k)}}  & \multicolumn{3}{c}{\textbf{Flickr30k (1k)}} \\  
    \cline{2-4} \cline{5-7} \addlinespace[1ex]
    & $\rho$-R@1$\uparrow$ & $\rho$-R@5$\uparrow$ & $\rho$-MRR@10$\uparrow$ & $\rho$-R@1$\uparrow$ & $\rho$-R@5$\uparrow$ & $\rho$-MRR@10$\uparrow$\\ 
    \midrule
    (BLIP, $\eta=1e-2$) & 61.0 & 65.7 & 72.3 & 54.7 & 55.0	& 63.9 \\ 
    (BLIP, $\eta=1e-3$) & 74.0 & 76.9 & 83.8 & 66.2 & 65.5	& 73.6\\ 
    (BLIP, $\eta=1e-4$) & 79.7 & 82.1 & 88.2 & 71.6 & 72.8	 & 79.3\\ 
    (BLIP, $\eta=1e-5$) & 81.2	 & 83.8	 & 89.2 & 74.3 & 74.0 & 81.1 \\ 
    \midrule
    (ALBEF, $\eta=1e-2$) & 64.4 & 68.7 & 75.5 & 57.7 & 57.0 & 67.5\\ 
    (ALBEF, $\eta=1e-3$) & 73.1 & 76.7 & 83.5 & 68.8 & 69.0 & 77.2\\ 
    (ALBEF, $\eta=1e-4$) & 78.1 & 80.7 & 87.2 & 73.2 & 74.6 & 81.3\\ 
    (ALBEF, $\eta=1e-5$) & 78.2 & 81.3 & 87.3 & 74.2 & 72.5 & 82.0\\ 
    \bottomrule
    \end{tabular}
\end{table}
This research question aims to analyze the faithfulness of sparsified models to their original dense models. We report in Table \ref{tab:dense-sparse-correlation} the Pearson correlation calculated for various effectiveness metrics of dense and sparsified queries. 
The results show that the correlation between sparsified and dense models is consistently positive and tends to increase as we relax the sparse regularization. Furthermore, as we consider higher cutoff values (R@1, R@5, MRR@10), the correlation tends to increase as the performance gap between the two systems narrows. Manually comparing the top-10 ranked images of the most differing queries, we find that while the two models rank top-10 images differently, there are a lot of common images (including the golden image) that look equally relevant to the query. Figure \ref{fig:images_overlap} shows that a high ratio (average: 70\%) of the top-10 images appear in both dense and sparse ranking lists. This analysis shows that the sparsified model is reasonably faithful to the dense model, suggesting that the sparse output terms could potentially be used for studying the semantics of dense vectors. 
\negskip



\subsection{Retrieval latency of dense and sparsified models}
We discussed the average FLOPs of sparsified models for retrieval efficiency. 
We now present query throughput and retrieval latency results in Table \ref{tab:latency}. Using Faiss~\cite{johnson2019billion} and PISA~\cite{MSMS2019, macavaney:sigir2022-pisa} on a single-threaded AMD Genoa 9654 CPU, the dense BLIP model with Faiss HNSW is exceptionally fast, outperforming D2S models with PISA. D2S models with query expansion (\textit{exp=c}) are slower due to high FLOPs and possibly LSR known limitations~\cite{mackenzie2021wacky}. Removing expansion terms (\textit{exp=0}) improves latency (FLOPs similar to DistilSPLADE~\cite{formal2021splade, splade_pp}) but is still approximately $30\times$ slower than dense retrieval. To balance efficiency and effectiveness of D2S, we propose using the inverted index with original query terms for retrieval, followed by re-scoring with expansion terms. With our simple iterative implementation, this approach proves effective, especially for retrieving fewer images per query. Surprisingly, indexing D2S models with Faiss HNSW competes well with PISA, particularly at higher cut-off values (100, 1000).
\begin{table}[t!]
    \centering
    \caption{Retrieval latency (CPU - 1 thread) of D2S models on 123k MSCOCO images.}
    \begin{tabular}{@{\extracolsep{2pt}}lccccccc@{}}
    \toprule
    \textbf{Model} &  & \multicolumn{3}{c}{\textbf{Throughput} (q/s)} & \multicolumn{3}{c}{\textbf{Latency} (ms)} \\ 
    \cmidrule{3-5} \cmidrule{6-8}
     & FLOPS & @10  & @100 & @1000 & @10 & @100 & @1000 \\ 
     \midrule
       Dense (BLIP, HNSW, Faiss) &  -  & 13277 & 9739 & 7447 & \phantom{00}0.08 & \phantom{00}0.10 & \phantom{00}0.14 \\
       \midrule
       D2S (BLIP, $\eta=1e-3$, exp=c, PISA)  & 11.5 & \phantom{0000}6  & \phantom{000}5 & \phantom{000}5 & 156.60 & 183.42 & 193.46 \\
       D2S (BLIP, $\eta=1e-3$, exp=0, PISA)  &  \phantom{0}2.8  & \phantom{00}449 & \phantom{0}284 & \phantom{0}160 & \phantom{00}2.23 & \phantom{00}3.52 & \phantom{00}6.25\\
       No Expansion $>>$ Expansion  & - & \phantom{00}369 & \phantom{0}120 & \phantom{00}18 & \phantom{00}2.70 & \phantom{00}8.31 & \phantom{0}54.05  \\ 
       \midrule
       D2S (BLIP, $\eta=1e-5$, exp=c, PISA)  &  78.4 &  $<$1 & $<$1 & $<$1  & $>$300 & $>$600 & $>$700 \\ 
       D2S (BLIP, $\eta=1e-5$, exp=0, PISA)  & \phantom{0}3.2  &  \phantom{00}230 & \phantom{0}146 & \phantom{00}90 & \phantom{00}4.34 &  \phantom{00}6.85  & \phantom{0}11.04  \\
       No Expansion $>>$ Expansion  & - & \phantom{00}189 & \phantom{00}70 & \phantom{00}11 & \phantom{00}5.30 & \phantom{0}14.37 & \phantom{0}86.66 \\
       \midrule
        D2S (BLIP, HNSW, Faiss) & - & \phantom{00}262 & \phantom{0}262 & \phantom{0}256 & \phantom{00}3.82 & \phantom{00}3.82 & \phantom{00}3.90\\
        \bottomrule
    \end{tabular}
    \label{tab:latency}
\end{table}


\negskip
\section{Conclusion}
\label{sec:conclusion}

\negskip
We have focused on the problem of efficiently transforming a pretrained dense retrieval model into a sparse model. 
We show that training a projection layer on top of dense vectors with the standard contrastive learning technique leads to the problems of dimension co-activation and semantic deviation. To mitigate these issues, we propose a training algorithm that uses a Bernoulli random variable to control the term expansion. Our experiments show that our \ourmodel{} sparsified model trained with the proposed algorithm suffers less from those issues. In addition, our sparsified models perform competitively to the state-of-the-art multi-modal LSR, while being faithful to the original dense models. 


\header{Acknowledgments}
We thank our anonymous reviewers for their valuable feedback.
This research was supported by
Ahold Delhaize, 
the Hybrid Intelligence Center, a 10-year program funded by the Dutch Ministry of Education, Culture and Science through the Netherlands Organisation for Scientific Research, 
project LESSEN with project number NWA.1389.20.183 of the research program NWA ORC 2020/21, which is (partly) financed by the Dutch Research Council (NWO),
project IDEAS with project number VI.Vidi.223.166 of the NWO Talent Programme,
which is (partly) financed by the Dutch Research Council (NWO),
and
the FINDHR (Fairness and Intersectional Non-Discrimination in Human Recommendation) project that received funding from the European Union’s Horizon Europe research and innovation program under grant agreement No 101070212.
All content represents the opinion of the authors, which is not necessarily shared or endorsed by their respective employers and/or sponsors.

\bibliographystyle{splncs04nat}
\bibliography{bibliography}

\end{document}